\documentclass{article}
\usepackage{amsfonts,amssymb,amsmath, epsfig}
\usepackage[american]{babel}
\usepackage{graphicx}

\newcommand{\bbi}{\mathbf{i}}
\newcommand{\bj}{\mathbf{j}}
\newcommand{\bn}{\mathbf{n}}

\newcommand{\no}{\noindent}

\newcommand{\ve}{\varepsilon}

\newcommand{\bnu}{\boldsymbol{\nu}}
\newcommand{\bxi}{\boldsymbol{\xi}}
\newcommand{\bomega}{\boldsymbol{\omega}}
\newcommand{\btau}{\boldsymbol{\tau}}

\def\pb{\, .}

\def\vb{\, ,}
\def\va{\,\raise 2pt\hbox{,}}

\def\bv{\boldsymbol{v}}

\def\bx{{\bf x}}

\def\bI{{\bf I}}
\def\bn{{\bf n}}

\def\a{{\alpha}}
\def\b{{\beta}}

\font\dc=cmbxti10 %slanted boldface for titles of "definition"

\begin{document}

\title{First Order Models of Human Crowds with Behavioral-Social Dynamics}

\author{Nicola Bellomo\\
\small SiTI - Istituto Superiore sui Sistemi Territoriali per l'Innovazione\\
\small Via Pier Carlo Boggio 61, 10138 Torino, Italy\\       
\small \texttt{nicola.bellomo@polito.it}\\      
\and
Stefano Berrone, \quad Livio Gibelli, \quad Alexandre Pieri\\
\small Department of Mathematical Sciences, Politecnico di Torino \\[-0.8ex]
\small Corso Duca degli Abruzzi 24, 10129 Torino, Italy \\
\small \texttt{stefano.berrone@polito.it} \\
\small \texttt{livio.gibelli@polito.it} \\
\small \texttt{alex.b.pieri@gmail.com>}}

\maketitle

\begin{abstract}
This paper presents a new approach to  behavioral-social dynamics of human crowds.
First order models are derived based on mass conservation at the macroscopic scale, while methods of the kinetic theory are used to
model the
decisional process by which walking individuals select their velocity direction.
Crowd heterogeneity is modeled by dividing the whole system into subsystems identified by different features.
The passage from one subsystem to the other is induced by interactions. 
It is shown how heterogeneous individual behaviors can modify the collective dynamics, as well as how local unusual behaviors can propagate in the 
crowd. The paper also proposes a system approach to the modeling of the dynamics in complex venues, where individuals move through
areas with different features. \\

\end{abstract}

%%%%%%%%%%%%%%%%%%%%%%%%%%%%%%%%%%%%%%%%%%%%%%%%%%%%%%%%%%%%%%%%%%%
%%%%%%%%%%%%%%%%%%%%%%%%%%%%%%%%%%%%%%%%%%%%%%%%%%%%%%%%%%%%%%%%%%%

\section{Plan of the Paper}

The modeling of  crowd dynamics can be developed, see~\cite{[BPT12]} and the book~\cite{[CPT14]}, at the three  scales,
namely \textit{microscopic} (individual based), \textit{macroscopic} (corresponding to the dynamics of mean averaged quantities), and to the intermediate   \textit{mesoscopic}, corresponding to the dynamics of a probability distribution function over the microscopic scale state of individuals. The latter approach is such that interactions are modeled at the micro-scale,  while mean quantities, such as local number density and linear momentum, are obtained by velocity weighted moments of the aforesaid probability distribution.

A critical analysis of the advantages and drawbacks of the different scales  selected for the modeling approach are discussed in the review
paper~\cite{[BD11]}, where it is stated that the present state of the art does not yet allow well defined hallmarks to support an optimal choice.
Therefore, due to these reasonings, a deep investigation has been developed to understand the complex dynamics at the microscopic scale, see,
among others,~\cite{[DAM13]} and~\cite{[HEL01]}.
The result can contribute to implement both meso-scale models~\cite{[BBK13]} or hybrid models~\cite{[ACK15]}, where the state of the system is
defined in probability over the velocity direction and deterministically over the velocity. A deep analysis of individual based interactions can contribute to derive, as we shall see, also hydrodynamic models \cite{[CC08],[HUG03]}.

Macroscopic, hydrodynamic, models are of great interest in that they are far less computationally demanding than those at  the other two scales.
This requirement is particularly important when dealing with complex flows such as coupling pedestrian flows to vehicular traffic
networks~\cite{[BKK14]}.
 However, their main drawback is that the heterogeneous behavior of walkers gets lost in the averaging process needed to derive models,
 which ends up with hiding this important feature. An additional difficulty, well documented in the paper by Hughes~\cite{[HUG03]}, consists in modeling the process by which
walkers select their velocity, namely direction and speed, in a crowd.

In general, an important issue to be taken into account in the modeling approach is the  nonlocal feature of interactions, as walkers are not classical particles and  modify their velocity before encountering a wall.

 Finally, let us stress that challenging problem to be considered is the propagation of anomalous behaviors, which might be induced by panic conditions \cite{[HFV00],[HJ09],[HJA07]}, that can induce large deviations in the collective dynamics. Due to all aforementioned motivation the term \textit{social dynamics} has been introduced in \cite{[BG15]} to enhance the heterogeneous behavior of walkers, who might modify their strategy induced by interaction not only with other individuals, but also with the specific features of the environment where they walk. The interested reader can find in \cite{[EFT12]} a valuable investigation on hyperbolic equations generated by nonlocal interactions, see also \cite{[BE14]}.

The present paper aims at tackling the aforesaid drawbacks in the case of first order models. These are simply obtained by mass conservation equations, which involve local density and mean velocity closed by phenomenological models relating the local mean velocity to local density distribution. Although this approach is a simple way of looking at the dynamics, substantial developments are needed with respect to the existing literature,  to achieve a realistic modeling of the decision process by which individuals select the velocity direction and adjust their velocity to the local density conditions, as well as the heterogeneous behavior of the crowd. Moreover, this paper proposes a systems approach to model the dynamics in complex venues, where walkers move through ares with different geometrical and qualitative features.

The contents of the paper are presented in four sections. In detail, Section 2 defines the mathematical structure underlying the modeling approach, which consists in a system of mass conservation equations for a crowd subdivided into various populations, which can be called after \cite{[BKS13]}, functional subsystems. This  structure acts as a background paradigm for the derivation of specific models. Three structures are derived corresponding to a homogeneous population, an heterogeneous populations, but without social dynamics, and a population where social dynamics is accounted for. Section 3 shows how specific models can be derived according to the aforesaid structure. More specifically, three classes of models are proposed corresponding to the hierarchy of models defined in the preceding section. Section 4 presents some perspective  ideas focusing on research plans, with more detail, the following topics are selected according to the authors' bias: modeling the heterogeneous  velocity distribution, propagation of anomalous behaviors, as well as an introduction to a systems approach to model the dynamics in complex environments, where nonlocal interactions can play an important role in the passage through different areas.

 %%%%%%%%%%%%%%%%%%%%%%%%%%%%%%%%%%%%%%%%%%%%%%%%%%%%%%%%%%%%%%%%%%%
 %%%%%%%%%%%%%%%%%%%%%%%%%%%%%%%%%%%%%%%%%%%%%%%%%%%%%%%%%%%%%%%%%%%

\section{Mathematical Structures}

Let us consider the dynamics of a crowd  in a domain $\Sigma$, which includes internal obstacles and inlet/outlet segments on the boundary $\partial \Sigma$.
This section searches for appropriate mathematical structures, which can provide the conceptual basis for the derivation of first order macroscopic models. It is assumed that the state of the system is described by local density and mean velocity to be viewed as
dependent variables of the differential system. However, since we deal with first order models, it is also necessary looking for a functional relation linking the mean velocity to the density.

Dimensionless quantities are used according to the following definitions:
\begin{itemize}
\item $\rho$ is the ratio between the number density $n$ of individuals per unit area and the number $n_M$ corresponding to the highest admissible
      packing density, namely $\rho := \frac{n}{n_M}$.
\vskip.2cm
\item $\bxi$ is the dimensionless local mean velocity obtained by referring the dimensional velocity $\bv$ to the highest limit of the mean velocity
      $v_M$, which can be reached by a walker in a low density limit in optimal environmental conditions, namely
      $\bxi := \frac{\bv}{v_M}$.
\end{itemize}

Density and velocity depend on time and space coordinates, namely $\rho = \rho(t,\bx) =: \rho (t,x,y)$ and $\bxi =  \bxi(t,\bx) =: \bxi (t,x,y)$.
The independent variables also set by dimensionless quantities by dividing $x$ and $y$ by a characteristic dimension $\ell$ of $\Sigma$ and time by $\frac{\ell}{v_M}$. Moreover, following \cite{[BG15]},  we introduce a number of dimensionless parameters that account for some specific features of the crowd already discussed in previous papers based on the kinetic theory approach:
\begin{itemize}
\item $\alpha \in [0,1]$ models the quality of the environment where $\alpha =1$ stands for the optimal quality of the environment, which allows to reach high velocity, while  $\alpha =0$ stands for the worst quality, which prevents the motion;

\vskip.2cm \item $\beta \in [0,1]$ models the attraction toward the direction of the highest density gradient, where $\beta =1$ stands for highest attraction when all individuals follow what the other do, while $\beta =0$ stands for the highest search of the less congested areas.
\end{itemize}
These parameters are used in this section only at a formal level, while their physical meaning will be made precise in the next section devoted to derivation of specific models.

Three levels of models are considered corresponding, respectively,  to a crowd with  homogeneous  walking ability, a crowd subdivided into different populations featured by different walking abilities, and finally a social crowd where interactions can modify social behaviors, that can also include anomalous ones.

\vskip.2cm \noindent $\bullet$ {\dc Homogeneous crowd:} Let us first consider the derivation of the mass conservation equation for an homogeneous crowd,  where all individuals have the same walking ability. Classically, the said equation writes as follows:
\begin{equation}\label{massc}
\partial_t \rho + \nabla_\bx \left( \rho\, \bxi \right) = 0\pb
\end{equation}

The closure of the equation can be obtained by modeling the dependence of $\bxi$ on $\rho$ by a phenomenological relation of the type $\bxi = \bxi[\rho](\alpha, \beta)$, so that the conservation equation formally writes as follows:
\begin{equation}\label{structureI}
\partial_t \rho + \nabla_\bx \left( \rho \, \bxi[\rho](\alpha, \beta) \right) = 0\vb
\end{equation}
where square brackets  denote that functional, rather than functions, relations can be used to link the local mean velocity to the local density. For instance, walkers select their preferred directions based not only on the local density, but also on density gradients. Therefore, specific models can be obtained by some heuristic interpretations of physical reality leading to $\bxi = \bxi[\rho]$ and inserting such model into Eq.(\ref{structureI}), which is the mathematical structure to be used for the modeling.

\vskip.2cm \noindent $\bullet$  {\dc Heterogeneous crowd:} Let us now consider a more general framework, where walkers are subdivided into a number $n$ of populations, the labeled by the subscript $i$, corresponding to different levels of expressing their walking abilities. Therefore, the state of the system is defined by a set of dimensionless number densities
\begin{equation}\label{densities}
 \rho_i = \rho_i(t,\bx) \vb \qquad i= 1, \ldots, n \vb \qquad \rho(t,\bx) = \sum_{i=1}^n \rho_i(t,\bx)\vb
\end{equation}
 The subscripts correspond to a discrete  variable, modeling the walking ability, with values corresponding to the the subscripts $i = 1, \ldots, n$, being
 $i=1$ and $i=n$, respectively,  to the lowest and highest motility.

The new structure simply needs the following modification:
\begin{equation}\label{structureII}
\partial_t \rho_i +  \,\nabla_\bx \left( \rho_i \bxi_i[\rho](\alpha,\beta) \right) = 0, \qquad i= 1, \ldots, n \vb
\end{equation}
where the modeling of the mean velocity differs for each population $\bxi_i = \bxi_i[\rho]$. Therefore, the structure consists in a
system on $n$ nonlinear PDEs equations.

\vskip.2cm \noindent $\bullet$  {\dc Social crowd:} Let us finally consider a dynamics, where each individual expresses a certain \textit{social variable}, for instance (but not only)  \textit{panic}, with values in the interval $[0,1]$, where the extremes denote the lowest and highest level of social expression corresponding to  such variable. Additional notations and a substantial development of the structure Eq.(\ref{structureII}) are  needed.
\begin{enumerate}

\item A social variable, say $w$, is introduced with discrete values $j= 1, \ldots, m$, where $j=1$ denotes the lowest level of expression, e.g.
absence of panic $w=0$, while $j=m$ denotes the highest admissible value $w=1$.

\vskip.1cm \item The overall system is subdivided into $N = n \times m$ populations labeled by the subscript $ij$.

\vskip.1cm \item The density in each population is denoted by $\rho_{ij}(t, \bx)$, therefore the total density is
$$
\rho(t, \bx) = \sum_{i=1}^n \, \sum_{j=1}^m\, \rho_{ij}(t, \bx).
$$

\vskip.1cm \item A source term $S_{ij}[\rho]$ models the transition from one functional subsystem to the other due to social interactions. The formal expression of such term is as follows:
\begin{equation}\label{source}
 S_{ij} = S_{ij}[\rho](\alpha, \beta, \gamma)\vb \quad \hbox{with} \quad \sum_{i=1}^n \,\sum_{j=1}^m S_{ij}[\rho](\alpha, \beta, \gamma) =0 \vb
\end{equation}
where $\gamma$ is a parameter triggering the transition into the $ij$-state, which depends on the specific type of social exchange.
\end{enumerate}

The mathematical structure now consists in a system of $n \times m$ nonlinear PDEs as follows:
\begin{equation}\label{structureIII}
\partial_t \rho_{ij} +  \,\nabla_\bx \left( \rho_{ij} \bxi_{ij}[\rho](\alpha,\beta) \right) =
S_{ij}[\rho](\alpha,\beta,\gamma) \vb \quad i= 1, \ldots, n \quad j= 1, \ldots, m\vb
\end{equation}
and where the modeling of the mean velocities $\bxi_{ij} = \bxi_{ij}[\rho]$ differ for each $ij$-population.

It is useful, for sake of completeness,  considering also the case of a crowd with a homogeneous mobility, but heterogeneous social state. Such a structure can be useful to study the propagation of anomalous behaviors. A simplification of Eq.~(\ref{structureIII}) yields:
\begin{equation}\label{structureIV}
\partial_t \rho_{j} +  \,\nabla_\bx \left( \rho_{j} \bxi_{j}[\rho](\alpha,\beta) \right) =
S_{j}[\rho](\alpha,\beta,\gamma) \vb \qquad j= 1, \ldots, m\pb
\end{equation}

\vskip.2cm \no \textbf{Remark 2.1} \textit{Structures defined by Eqs.~(\ref{structureI}),--(\ref{structureII}) and (2.6) constitute the underlying framework
to derive specific models. In all cases, the derivation depends on the environment, where the crowd moves, not only its quality, but also its shape. In fact, the model requires implementing nonlocal interaction with the walls. Moreover, transferring the structure (\ref{structureIII}) into a model requires a detailed analysis of the social dynamics.}

\vskip.2cm \no  \textbf{Remark 2.2} \textit{Specific models, where some social dynamics is taken into account, need technically a specification of the kind of social information exchanged by walkers.}

\vskip.2cm  \no  \textbf{Remark 2.3} \textit{The formal structures proposed in this section need three types of parameters, namely $\alpha$, $\beta$, and $\gamma$, corresponding to three specific features of the dynamics. Their practical implementation into models will be clarified in the next section.}

%%%%%%%%%%%%%%%%%%%%%%%%%%%%%%%%%%%%%%%%%%%%%%%%%%%%%%%%%%%%%%%%%%%
%%%%%%%%%%%%%%%%%%%%%%%%%%%%%%%%%%%%%%%%%%%%%%%%%%%%%%%%%%%%%%%%%%%

\section{Derivation of Models}

 Let us now consider the derivation of specific models consistent with the frameworks presented in the preceding section. Therefore, three different classes of models are derived in the following subsections according to conceivable requirements of  applications. More in detail, we consider both homogeneous and  heterogeneous populations in absence of social interactions, and an heterogeneous population when social exchanges take place so that individuals are allowed to shift from one subpopulation to the other. The study of panic conditions or, more in general, propagation of anomalous behaviors can be treated in the latter case. These contents are presented in the next subsection, while the last one proposes a critical analysis looking ahead to research perspectives.

\subsection{Modeling an homogeneous crowds}

Let us consider the simple case of an homogeneous crowd, where all individuals behave in the same manner. Derivation of models requires simply to describe
analytically the dependence of $\bxi$ on the local density distribution. Polar coordinates are used for the velocity variable, so that
\begin{equation}
\bxi = \xi\, (\cos \theta \, \bbi + \sin \theta \, \bj) := \xi \, \bomega ,
\end{equation}
where $\theta$ is the angle between $\bxi$ and the $x$-axis of an orthogonal system in the plane,  $\bbi$ and $\bj$ are the unit vector of the aforesaid axes, $\xi$ is the velocity modulus of $\bxi$ occasionally called \textit{speed}.

The idea that pedestrians adjust their dynamics according to a decision process, which can modeled by theoretical tools of game theory \cite{[NOW06]}, was already introduced in \cite{[BB11]} and subsequently developed in a sequel of papers \cite{[ACK15],[BBK13],[BG15]}.

In detail, the modeling proposed in this paper takes advantage of the approach to  individual based interaction proposed in \cite{[BG15]}, where the  decision process is supposed to act according to the following sequence:

\begin{enumerate}
\vskip.2cm \item Walkers move along the direction $\bomega$, forming an angle  $\theta$ with respect to the $x$-axis, selected according to a decision process which account the following trends: search of the exit, avoid walls, search of less congested areas, and attraction toward the density gradients. Details are given in the following.
\vskip.2cm \item Once the direction $\bomega$ has been selected, pedestrians adjust their velocity modulus  to the so-called local \textit{perceived density} $\rho^*_\theta$ along $\bomega$. This quantity is defined, according to \cite{[BBNS14]} as follows:
\begin{equation}\label{perceived}
\rho^*_\theta  = \rho^*_\theta [\rho]= \rho + \frac{\partial_\theta \rho}{\sqrt{1 + (\partial_\theta \rho)^2}}\,\big[(1- \rho)\, H(\partial_\theta \rho) + \rho \, H(- \partial_\theta \rho)\big]\va
\end{equation}
where $\partial_\theta$ denotes the derivative along the direction $\theta$, while  $H(\cdot)$ is the Heaviside  function $H(\cdot \geq 0) = 1$, and $H(\cdot < 0) = 0$. Therefore, positive gradients increase the perceived density up to the limit $\rho = 1$, while negative gradients decrease it down to the limit $\rho = 0$ in a way that
\begin{equation}\label{limits}
\partial_\theta \rho \to \infty \Rightarrow \rho^*_\theta \to 1\vb \quad \partial_\theta  \rho = 0 \Rightarrow \rho^*_\theta = \rho \vb \quad
\partial_\theta  \rho \to - \infty \Rightarrow \rho^*_\theta \to 0\pb
\end{equation}

\vskip.2cm \item  The relation  $\xi = \xi(\rho^*_\theta;\a)$ between $\xi$ and $\rho^*_\theta$ depends on the quality of the environment. In the attempt of reproducing empirical data \cite{[DH03],[MHG09],[MT11],[SS11],[SSKB06]}, is modeled by a polynomial of $\rho^*_\theta$ fulfilling the following constraints:  $\xi(0) = \a;\, \xi'(0) = 0;\, \xi(1) = \a;\, \xi'(1) = 0$, where prime denotes derivative with respect to $\rho^*_\theta$.
\end{enumerate}

The first step of the above process leads, according to \cite{[BG15]} to the following expression of the preferred direction:
\begin{equation}\label{theta}
\bomega_F (\rho, \bx; \beta) =
\frac{(1-\rho) \bnu_T + \rho \left[ \beta \bnu_S + (1-\beta) \nu_V \right]}
               {\left\| (1-\rho) \bnu_T + \rho \left[ \beta \bnu_S + (1-\beta) \bnu_V \right] \right\|}\va
\end{equation}
where
\begin{equation}
      \bnu_S = -  \bnu_V = \frac{\nabla \rho}{|| \nabla \rho ||}\va
\end{equation}
while the parameter $\beta$ may account for panic conditions \cite{[BG15]}.
Hence, $\cos{\theta} = \bomega \cdot \bbi,  \sin{\theta} = \bomega \cdot \bj$.

Neglecting the influence of the walls, the second and third step yield the following polynomial expression:
\begin{equation}\label{speed}
\xi_F = \xi_F[\rho] = \a(1 - 3\, {\rho^*_\theta}^2 + 2\, {\rho^*_\theta}^3)[\rho].
\end{equation}

These two quantities can be inserted into the framework Eq.~(\ref{massc}), which writes as follows:
\begin{equation}\label{massc2}
\partial_t \rho + \nabla_\bx \left(\a\, \rho (1 - 3\, {\rho^*_\theta}^2 + 2\, {\rho^*_\theta}^3)\, \bomega_F(\rho,\bx;\b)\right) = 0\pb
\end{equation}

The role of the nonlocal interactions with the wall and obstacles can be modeled by adapting the hallmarks of \cite{[BG15]} to this present case.
The strategy pursued is that the aforementioned interactions, which replace the classical local boundary conditions, define a preferred walking
direction according to a two-steps procedure.
As a first step, walkers change the direction of motion to $\bomega_F$.
As a second step, walkers may further change the direction from $\bomega_F$ to $\bomega$ if two conditions occur:

\begin{enumerate}
 \item The  walker's distance from the wall, $d$, is within a given cutoff distance, $d_w$;
\vskip.2cm  \item the walker's velocity is directed toward a wall.
\end{enumerate}

In detail, if $\bxi \cdot \bomega_F <0$ and $d<d_w$ , the components of $\bomega_F$ are decomposed into the normal and tangential components,
respectively
\begin{equation}
\omega_1 (\rho, \bx; \beta)=(\bn \otimes \bn)  \bomega_F (\rho, \bx; \beta), \quad
\hbox{and} \quad
\omega_2 (\rho, \bx; \beta)=(\bI-\bn \otimes \bn)  \bomega_F (\rho, \bx; \beta),
\end{equation}
where $\bn$ is the unit vector orthogonal to the wall. Subsequently, it is assumed that
the velocity component normal to the wall decreases linearly
while approaching the wall and becomes naught in the limiting case of a pedestrian in contact with the wall, hence
\begin{equation}
\bomega (\rho, \bx; \beta) =  \frac{d}{d_w} \omega_1(\rho, \bx; \beta) \bn + \left(1-\frac{d^2}{d_w^2}\omega_1^2 (\rho, \bx; \beta) \right)^{1/2} {\btau} \vb
\end{equation}
where $\btau$ is the unit vector tangential to the wall.
The model is obtained by Eq.~(\ref{massc2}) simply by replacing $\bomega_F$ with $\bomega$.

\subsection{Heterogeneous crowd in absence of social dynamics}
The reference mathematical structure is now given by  Eq.~(\ref{structureII}), which consists in a system of $n$ PDEs. Hence, the modeling problem
consists in modeling the mean velocity $\bxi_i$ for each population of pedestrians, which, as an example, can correspond to slow, mean, and fast individuals.
Hereinafter, the model is presented  in the case of a system in absence of boundaries, referring the technical generalization to the approach proposed in the preceding subsection. By assuming that pedestrians exploit the quality of the environment depending also on the quality of their walking ability, the following model is obtained:
\begin{equation}
\xi_i = \xi_i[\rho] =  \a\, \frac{i}{n} \, (1 - 3\, {\rho^*_\theta}^2 + 2 \, {\rho^*_\theta}^3)[\rho]
\end{equation}
which inserted into the structure yields:
\begin{equation}\label{mass-mix}
\partial_t \rho_i +  \,\nabla_\bx  \,\left( \rho_i\, \a \, \frac{i}{n} \, (1 - 3\, {\rho^*_\theta}^2 + 2\, {\rho^*_\theta}^3)[\rho] \, \bomega_i (\rho. \bx;\beta)
 \right) = 0\vb \quad i= 1, \ldots, n \vb
\end{equation}
where the preferred direction $\bomega_i$ is now computed for all components of the fluid viewed as a mixture.

\subsection{Modeling crowds with social dynamics}
Let us now consider the dynamics of individuals according to the general framework, which introduces a discrete social variable. Specific models can be obtained by a detailed description of the source terms $S_{ij}$. Therefore, let us   consider the modeling of transitions, induced by interactions, across functional subsystems to the other. These can be viewed as net flows, namely the inflows  into the $j$-subsystem from  the $(i-1)$- and $(i+1)$-subsystems, and minus the outflow from the $j$-subsystem into  the $(i-1)$ and $(i+1)$ subsystems. A formal expression is as follows:
\begin{eqnarray}
S_{ij}[\rho] = S^+_{ij}[\rho] - S^-_{ij}[\rho]&=& (\gamma^+_j \rho_{i(j-1)} \rho_{ij} - \gamma^-_i \rho_{i(j+1)} \rho_{ij})\nonumber \\ &-& (\gamma^+_i \rho_{i(j+1)} \rho_{ij} - \gamma^-_i \rho_{i(j-1)} \rho_{ij}),
\end{eqnarray}
where $\gamma^+_i$ and $\gamma^-_i$ are positive constants modeling, for each $i$, the flow from the lower to the upper state and vice versa and where the flow depends on the density of the state origin of the flow and that attracting it. Letting now $\gamma_i = \gamma^+_i - \gamma^-_i$, which is a parameter with positive or negative values, yields
\begin{equation}\label{flow-social}
S_{ij}[\rho] = S^+_{ij}[\rho] - S^-_{ij}[\rho] = \gamma_i \, (\rho_{i(j-1)} - \rho_{i(j+1)}).
\end{equation}
Therefore
\begin{equation}
\ve > 0: \quad   \rho_{i-1} \leq \rho_{i+1} \quad  \Rightarrow \quad S_{ij} \leq 0, \qquad \rho_{i-1} \geq \rho_{i+1} \quad  \Rightarrow \quad S_{ij} \geq 0.
\end{equation}

\vskip.2cm \no \textbf{Remark 3.1} \textit{This result can be specifically related to modeling panic, which is a situation such that initially $\rho_{i(j-1)}  \cong \rho_{ij}$, while increasing of panic shifts individuals to high values of $j$, namely $\rho_{i(j-1)}  < \rho_{ij}$. This topic will be further discussed in the last section.}

%%%%%%%%%%%%%%%%%%%%%%%%%%%%%%%%%%%%%%%%%%%%%%%%%%%%%%%%%%%%%%%%%%%%
%%%%%%%%%%%%%%%%%%%%%%%%%%%%%%%%%%%%%%%%%%%%%%%%%%%%%%%%%%%%%%%%%%%%

\section{From a Critical Analysis to Perspectives}

Three classes of models have been proposed in preceding sections, in sequence, corresponding to: A population with homogeneous behavior, an heterogeneous walking
ability and strategy, and to a dynamics which involves social behaviors, specifically panic.
The advantage of the approach proposed in this paper mainly relies on the simplicity of the mathematical structures used for the modeling  mass conservation closed by heuristic models linking mean velocity to local density distribution. This strategy leads to models which present a lower computational complexity with respect to the kinetic type approach \cite{[BG15]}. On the other hand some descriptive ability is lost in the averaging process induced by the hydrodynamic approach.

One of the aims of this paper has been developing a new approach suitable to include, as far as possible, some features of the heterogeneity,
which appears in crowd dynamics. This final section aims at understanding how far the mathematical structures proposed in the preceding sections can be further developed to model additional complexity features, heterogeneity in particular, already described in  \cite{[BG15]}. This development takes advantage of Gromov's hint toward a deep investigation of mathematical structures with the aim of discovering their richness \cite{[GRO12]}.

The contents focus on the following topics selected, among various ones, according to the authors bias: Additional reasonings on modeling different types of social behaviors, modeling  heterogeneity in the velocity distribution, and finally a system approach to the dynamics in complex venues, where the passage to different types of areas needs to be carefully modeled by taking into account nonlocal interactions. These topics are treated in the next subsections, which although not formalized by mathematical equations, propose various hints to tackle the related  conceptual difficulties.

\subsection{Modeling different types of social behaviors}

The modeling proposed in Subsection 3.3 is focused on panic conditions. However, a variety of different types of exchanges can appear. The modeling
of interactions can be achieved by suitable development of theoretical tools of game theory \cite{[GIN09],[NOW06]}. A recent literature on the mathematics of social systems provides useful hints toward this approach, see among others \cite{[ABE08],[BHT13],[DL14],[DP14]}. In particular, the following types of social exchanges can be considered:

\begin{itemize}

\item \textit{Consensus dynamics:} When walkers approach, during interactions, their mutual social status, namely individuals with low values of the activity variable shift to the higher value, while those with an high value shift to the lower one.

\vskip.2cm  \item \textit{Competitive dynamics:} When individuals behave in the opposite way  and increase the distance between  their mutual social status, namely individuals with low values of the activity variable shift to a lower value, while those with an high value shift to an higher one.

\vskip.2cm  \item \textit{Leaders dynamics:} When the presence of leaders induces a different dynamics as leaders do not change their status, while the other individuals shift toward the status of the leaders.
\end{itemize}

A detailed formalization of this qualitative description leads to the modeling of the fluxes, which appear in Eq.~(2.6) and hence to mathematical models related
to the three different types of dynamics under consideration.

\subsection{Modeling heterogeneity in the velocity distribution}

The modeling approach proposed in this paper shows a drawback typical of all hydrodynamic models, namely the averaging process hides the heterogeneity features of the system. In this present paper, heterogeneity of individuals is taken into account by the subdivision into functional subsystem, but the heterogeneity in the velocity distribution is lost. This feature obliges one to use empirical data to refer the speed to the local density distribution.

The empirical relation defined in Eq.~(\ref{speed}) links the speed  to the parameter $\alpha$ and to a polynomial expression of the perceived density, where
 $\alpha$ has the meaning of highest observed speed within a representation which is one-dimensional due to the fact the velocity direction has been selected according to the decision process followed by walkers. A more refined approach might consider $\xi$ as the first order moment of a probability distribution, evolving in time according to  interactions at the micro-scale to be properly investigated.

This objective can be achieved by taking advantage of the modeling approach proposed in \cite{[BDF12]} for vehicular traffic according to the following hallmarks:

\begin{itemize}

\item Walkers select their optimal velocity direction $\bomega$ according to the deterministic decision process proposed in the preceding section;

\vskip.2cm \item Speed is assumed to be given by a random variable $v(t;\alpha, \bx, \rho)$ depending locally on $\alpha$ and on the perceived density;

\vskip.2cm \item The variable $v$ is linked to a probability distribution $P(t,v|\alpha, \bx, \rho)$, whose dynamics is derived, as in  \cite{[BDF12]}, by individual based interactions depending on $\alpha$ and on the perceived density;

\vskip.2cm \item The mean speed $\xi$ is obtained as first moment of $P(t,v)$ for a system, where the overall dynamics is delivered by coupling the dynamics
of $\rho$ to that of  $P(t,v)$.

\end{itemize}

%%%%%%%%%%%%%%%%%%%%%%%%%%%%%%%%%%%%%%%%%%%%%%%%%%%%%%%%%%%%%%%%%%%
%%%%%%%%%%%%%%%%%%%%%%%%%%%%%%%%%%%%%%%%%%%%%%%%%%%%%%%%%%%%%%%%%%%

 \end{document}